\documentclass[journal=jacsat, manuscript=article, layout=onecolumn]{achemso}
\setkeys{acs}{keywords = true}
\usepackage[version=3]{mhchem} 
\usepackage[colorlinks = true, urlcolor = blue, linkcolor = blue, citecolor=blue]{hyperref}
\usepackage{amsmath,amssymb}
\usepackage{graphicx}
\usepackage{dcolumn}
\usepackage{bm}
\usepackage{color} 
\usepackage{multirow}
\usepackage{siunitx}
\newcommand{\onlinecite}[1]{\hspace{-1 ex} \nocite{#1}\citenum{#1}}

\author{Sejoong Kim}
\affiliation{University of Science and Technology (UST), Gajeong-ro 217, Daejeon 34113, Korea}
\alsoaffiliation{Korea Institute for Advanced Study, Hoegiro 85, Seoul 02455, Korea}
\email{sejoong@alum.mit.edu}

\title[An \textsf{achemso} demo]
  {DFT+$U$+$V$ Study of Magnetic Ordering in Single-Layer Pentahexoctite: Implications for Magnetic Device Platforms}

\keywords{pentahexoctite, 2D materials, carbon allotropes, magnetism, density functional theory, flat bands, Dirac semimetal}
\begin{document}







\begin{abstract}
In this work, we investigate the electronic and magnetic properties of single-layer pentahexoctite, a two-dimensional carbon allotrope patterned by pentagons, hexagons, and octagons. 
Using density functional theory (DFT) calculations incorporating on-site and inter-site Coulomb interactions, we find that type-II Dirac fermions are formed by a nearly flat band intersecting with a dispersive band at the Fermi level. 
We further examine the physical origin of nearly flat bands of pentahexoctite. 
Constructing \textit{ab initio} tight-binding Hamiltonian based on Wannier functions, we reveal that nearly flat bands of pentahexoctite originate from quantum-mechanical destructive interference. 
Remarkably, our DFT calculations including extended Hubbard interactions show that hole doping induces a ferrimagnetic phase transition driven by the enhanced density of states in the nearly flat band. 
This finding highlights that monolayer pentahexoctite is a promising candidate for pristine all-carbon magnetic materials to serve as a platform for future magnetic and spintronic devices.
\end{abstract}

\section{\label{sec:Intro}INTRODUCTION}
Graphene, a $sp^2$-hybridized hexagonal carbon lattice, has become a seminal two-dimensional (2D) material studied across diverse fields including physics, chemistry, and engineering~\cite{RevModPhys.81.109}. 
Its low-energy physics, governed by isotropic Dirac cones at the Fermi level, grants graphene intriguing properties like Klein tunneling~\cite{Katsnelson2006} and the anomalous integer Hall effect~\cite{Zhang2005, doi:10.1126/science.1137201}. 
Moreover, graphene's topological nature opens doors for understanding the quantum spin Hall effect~\cite{PhysRevLett.95.226801} and its generalization to topological insulators~\cite{PhysRevLett.98.106803}.  

Graphene has led to the exploration of diverse 2D materials including silicene~\cite{PhysRevLett.102.236804}, phosphorene~\cite{Carvalho2016}, transition metal dichalcogenides~\cite{Manzeli2017}, etc.
Research on 2D materials has also expanded to investigate novel lattice structures like van der Waals heterostructures~\cite{Geim2013} and twisted bilayer systems~\cite{Andrei2020, Cao2018, doi:10.1073/pnas.1108174108}. 
In addition, graphene inspired to explore various 2D carbon allotropes with $sp$, $sp^2$, and $sp^3$ hybridizations, such as graphyne~\cite{Hu2022}, graphdiyne~\cite{B922733D},  biphenylene~\cite{doi:10.1126/science.abg4509.Fan}, 
graphenylene~\cite{C2TC00006G},
pentagraphene~\cite{doi:10.1073/pnas.1416591112}, and pentaheptite~\cite{PhysRevB.53.R13303}, some of which have been experimentally synthesized~\cite{B922733D, Hu2022, doi:10.1126/science.abg4509.Fan}. 

Pentahexoctite, a theoretically predicted all-carbon planar material with pentagonal, hexagonal, and octagonal carbon rings~\cite{Sharma2014}, has been investigated for various properties including electronic, mechanical, optical, and charge transport properties~\cite{BRANDAO2022111686, PhysRevMaterials.2.085408, FERREIRA2022115468, BRANDAO2023104694}. 
However, many intriguing properties of pentahexoctite remain unexplored.
One interesting feature is that pentahexoctite exhibits a nearly flat band around the Fermi level~\cite{Sharma2014}. 
When pentahexoctite was initially proposed~\cite{Sharma2014}, this nearly flat band was reported, but a detailed analysis and its physical implications in the context of correlation physics are yet to be studied.  

Flat bands can occur in various lattice systems. 
Flat bands have been identified not only in artificial systems like optical lattices~\cite{PhysRevA.82.041402} but also in real materials such as Kagome metals~\cite{Kang2020_1}. 
In addition, it is also reported that flat bands exist in 2D carbon materials, for example, magic-angle twisted bilayer graphene (MATBG)~\cite{PhysRevLett.122.106405, PhysRevLett.106.126802} and biphenylene network (BPN)~\cite{doi:10.1021/acs.nanolett.2c00528.Son, PhysRevB.104.235422}, both of which have been experimentally realized~\cite{Cao2018, doi:10.1126/science.abg4509.Fan}. Flat bands, contributing to the diverging density of states, give rise to intriguing emergent physics, including superconductivity~\cite{Cao2018, Tian2023} and anomalous Landau levels~\cite{Rhim2020}.

Magnetism is another possible correlation-related physics induced by flat bands. Considering the picture of Stoner-type magnetism for itinerant electrons~\cite{Stoner1938}, it is possible that the interplay of enhanced density of states and exchange interactions can facilitate the emergence of magnetic ordering in the single-layer pentahexoctite. While graphene is a representative example of carbon-based magnetism, graphene magnetism typically requires specific conditions such as geometrical manipulations like zigzag nanoribbons~\cite{Son2006} and vacancies~\cite{PhysRevLett.93.187202}, hydrogenation~\cite{doi:10.1021/nl9020733}, or adatom introduction~\cite{PhysRevLett.91.017202}. However, pristine magnetic 2D materials are rare~\cite{GAO202111}. 
Thus it is worth searching for pristine 2D carbon-based material platforms for future spintronics and nanoelectronics applications. 
Considering the existence of the nearly flat band around the Fermi level, it is valuable to examine the possibility of magnetic ordering in monolayer pentahexoctite, potentially making it a promising candidate for pristine 2D magnetic carbon materials utilized as a platform for future magnetic and spintronic devices.

In this work, we investigate the formation of flat bands and their impact on magnetic ordering in monolayer pentahexoctite. Using the density functional theory (DFT) calculations~\cite{PhysRev_136_B864_1964, PhysRev_140_A1133_1965}, we analyze the electronic properties of monolayer pentahexoctite including band structures and Fermi surfaces, especially focusing on the location of the flat bands relative to the Fermi level. Moreover, we find that the flat band of pentahexoctite intersects with a dispersive band, thereby forming the type-II Dirac point~\cite{Nature.527.495.Soluyanov2015, PhysRevLett.115.265304}. 
This is in contrast to the type-I Dirac point of graphene. 
Constructing the \textit{ab initio} tight-binding model, we examine quantum-mechanical destructive interference paths contributing to the development of flat bands. Finally, employing extended Hubbard interactions and hole doping, we explore the possibility for magnetic transitions in monolayer pentahexoctite. 

\begin{figure}[t]
\begin{center}
\includegraphics[width=0.7\columnwidth, clip=true]{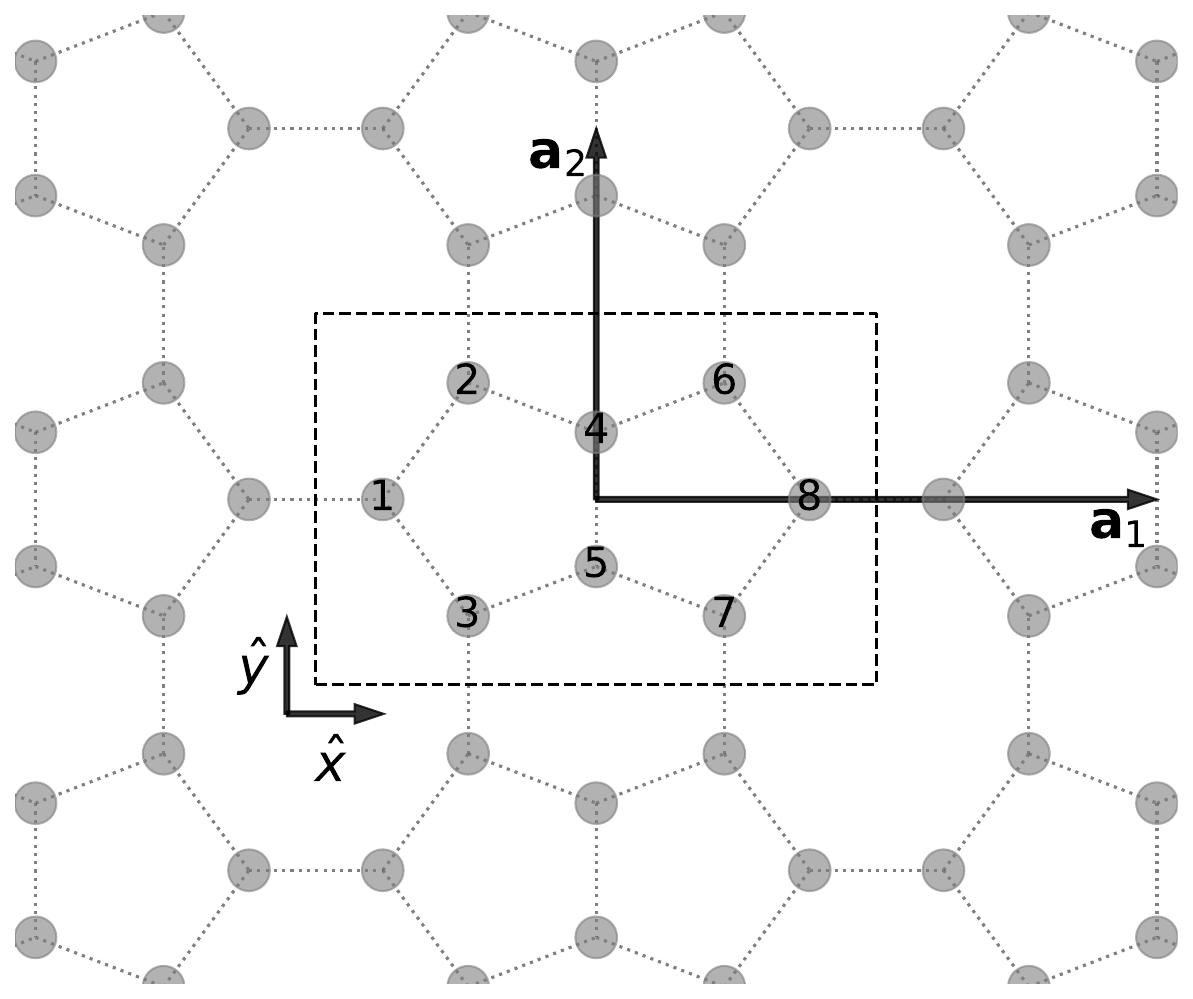} 
\end{center}
\caption{\label{fig:configuration} (Color online) (a) In-plane atomic configuration of pentahexoctite layer. The Wigner-Seitz cell is indicated by 
dashed lines. Inside the Wigner-Seitz cell, carbon atoms are numbered from 1 to 8. Atoms from 1 to 5 constitute the left pentagon, and atoms from 4 to 8 belong to the right pentagon. Lattice vectors $\mathbf{a}_{1}=a_{1} \hat{x}$ and $\mathbf{a}_{2}=a_{2} \hat{y}$ are indicated.}
\end{figure}
\section{\label{sec:DFT}COMPUTATIONAL DETAILS}
We perform DFT calculations by using \textsc{Quantum Espresso}~\cite{JPhys_CM_29_465901_2017} with the plane-wave (PW) basis, the PBE exchange-correlation functional~\cite{PhysRevLett_77_3865_1996} and norm-conserving pseudopotentials~\cite{PhysRevB_88_085117_2013}. 
For self-consistent calculations, we use $16\times16\times1$ $k$-point mesh, and the kinetic energy cutoff $100$ Ry. 
We take 15 $\textrm \AA$ as a vacuum size to prevent interactions between periodic image layers.
Furthermore, to investigate the effect of electron interactions, we use the DFT+$U$+$V$ method recently developed with \textsc{Quantum Espresso}, where on-site ($U$) and inter-site ($V$) Hubbard interactions are combined with DFT calculations~\cite{PhysRevResearch.2.043410}.
On-site and inter-site Hubbard interactions are self-consistently determined by using the Hartree-Fock formalism based on the pseudohybrid Hubbard density functional known as ACBN0~\cite{PhysRevResearch.2.043410, PhysRevB.102.155117, PhysRevX.5.011006}. 
We relax all atomic structures to ensure that all components of forces acting on atoms are below 0.051 eV/{\AA}.  
In this work, conventional DFT calculation with GGA functionals is denoted by DFT-GGA to distinguish against the DFT+$U$+$V$ method. 

\begin{table}[t]
\begin{tabular}{ c c c | c c c}
&&                              & DFT-GGA & DFT+$U$+$V$ &\\
\hline
&$a_{1}$& & 5.788   & 5.740       &\\
&$a_{2}$& & 3.828   & 3.863       &\\
&$d_{12}$&                      & 1.491   & 1.513       &\\
&$d_{24}$&                      & 1.417   & 1.400       &\\
&$d_{45}$&                      & 1.386   & 1.410       &\\
&$\theta_{124}$&               & 105.2$^{\circ}$ & 104.6$^{\circ}$ &\\
&$\theta_{245}$&                & 111.1$^{\circ}$ & 111.6$^{\circ}$ &\\
\end{tabular}
\caption{\label{table01} Atomic structure parameters of pentahexoctite monolayer optimized by DFT-GGA and DFT+$U$+$V$ calculations. $d_{ij}$ represents a bond length between carbon atoms $i$ and $j$. $\theta_{ijk}$ denotes an angle between bonds $\overline{ij}$ and $\overline{jk}$. Lattice constants $\left|\mathbf{a}_{i}\right|$ and bond lengths $d_{ij}$ are in a unit of angstrom ({\AA}), and angles $\theta_{ijk}$ are in a unit of degree ($^\circ$).}
\end{table}

\section{\label{sec:Band_Sturcture}BAND STRUCTURES}
\begin{figure}[t]
\begin{center}
\includegraphics[width=0.9\columnwidth, clip=true]{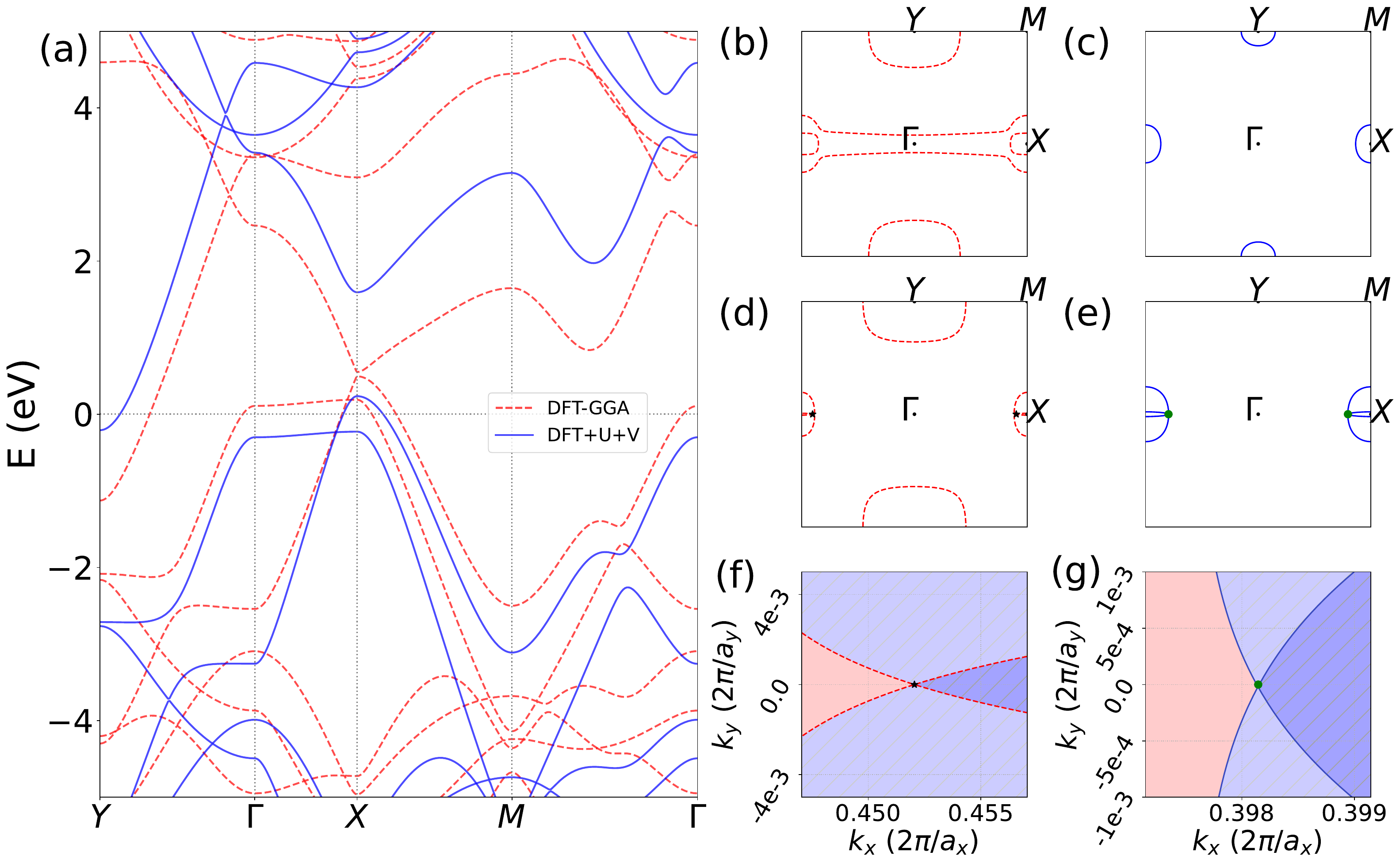} 
\end{center}
\caption{\label{fig:bands} (Color online) (a) Band structures of monolayer pentahexoctite calculated by DFT-GGA (red dashed lines) and by DFT+$U$+$V$ (blue solid lines). The Fermi energy $E_{F}$ is set to be zero. Fermi surfaces of DFT-GGA bands (red dashed lines). (c) Fermi surfaces of DFT+$U$+$V$ bands (blue solid lines). Fermi surfaces for (d) DFT-GGA bands and (e) DFT+$U$+$V$ bands when the Fermi energy is aligned to the Dirac point near $X$. The Fermi surfaces of DFT-GGA and DFT+$U$+$V$ in the vicinity of the type-II Dirac points are displayed in (f) and (g), respectively. The red area and blue hatched area indicate electron and hole regions, respectively. The dark blue and light blue denote hole regions of the flat band and the steep band, which constitute the type-II Dirac point. Dirac points of DFT-GGA bands and DFT+$U$+$V$ ones are denoted by black stars and green circles, respectively.}
\end{figure}

\begin{figure}[t]
\begin{center}
\includegraphics[width=1.0\columnwidth, clip=true]{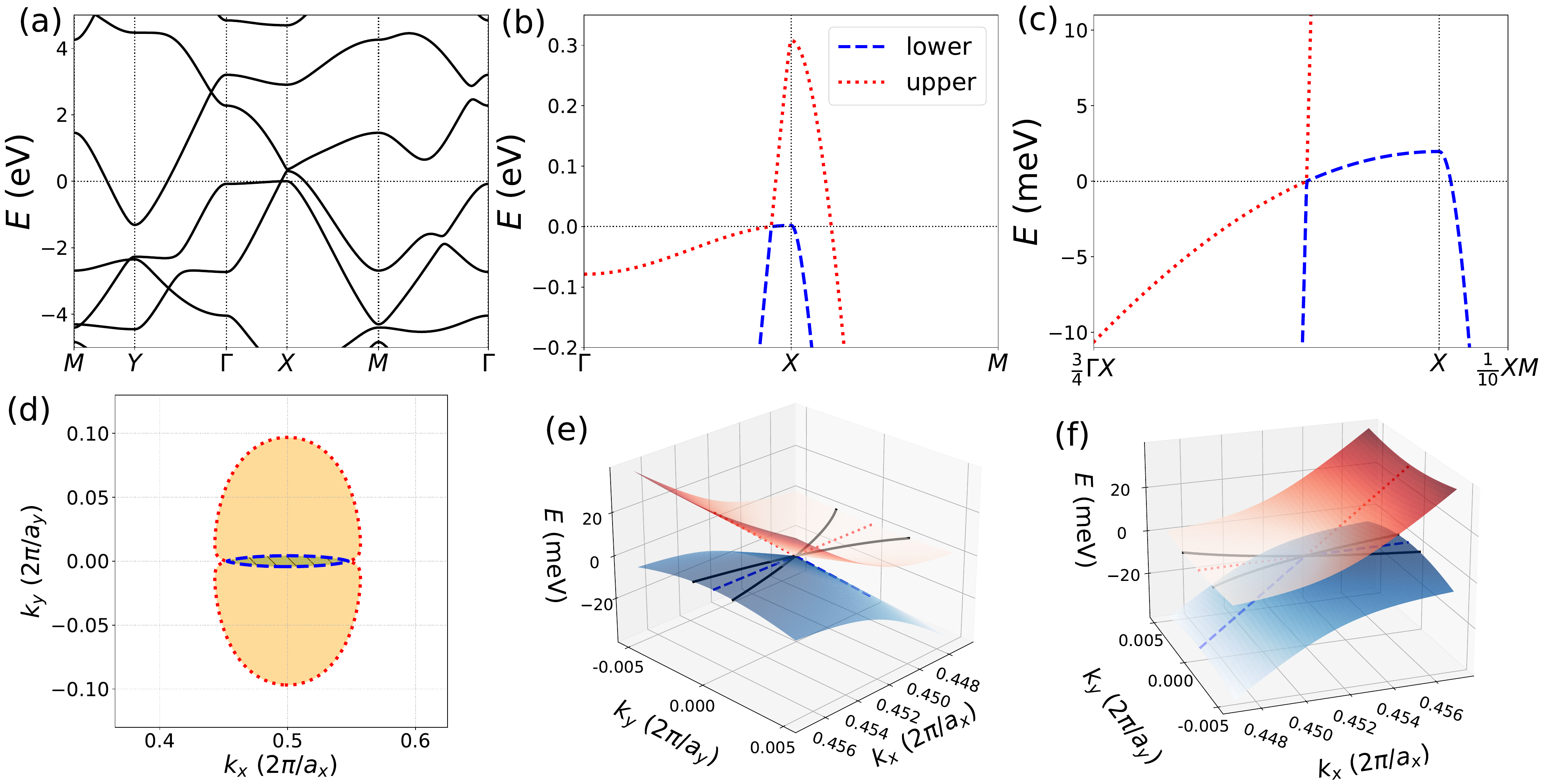} 
\end{center}
\caption{\label{fig:TypeII} (Color online) The type-II Dirac point of pentahexoctite. (a) Band structures of monolayer pentahexoctite calculated by DFT-GGA when the Fermi energy is aligned to the Dirac point. The Fermi energy is set to be zero. (b) Two energy bands constituting the type-II Dirac point at symmetric lines $\overline{\Gamma X M}$. The upper band and the lower one are denoted by red dotted lines and blue dashed ones, respectively. (c) The upper and lower bands around the type-II Dirac point. (d) The orange area and red dotted lines represent hole pockets and Fermi surfaces of the upper band. The blue hatched area and blue dashed lines indicate hole pockets and Fermi surfaces of the lower band. (e) and (f) 3D visualization of the upper (red) and the lower (blue) bands in the vicinity of the type-II Dirac point. Black solid lines represent Fermi surfaces. Red dotted lines (blue dashed lines) denote the upper (the lower) energy bands at $\overline{\Gamma M}$.}
\end{figure}

Figure~\ref{fig:configuration} illustrates the atomic configuration of pentahexoctite crystal, in which pentagons, hexagons, and octagons are periodically patterned. 
Here the unit cell comprises two pentagons sharing two carbon atoms. 
This atomic structure of pentahexoctite was theoretically proposed by Ref.~\onlinecite{Sharma2014}, where its energy is compared with that of the existing 2D carbon allotropes. 
It is reported that the energy of pentahexoctite is relatively higher than graphene by about 0.35 eV/atom, but it is comparable to that of pentaheptite. 
It is remarkable that the energy of pentahexoctite is lower than that of graphyne, graphdiyne, and graphenylene. 
Considering that these higher-energy structures have been synthesized, pentahexoctite could be experimentally synthesized in the future. 
In fact, Ref.~\onlinecite{Sharma2014} suggested the irradiation of the graphene by an electron beam in a controlled fashion as a possible realization method.

We first obtain fully relaxed structures of monolayer pentahexoctite using DFT-GGA and DFT+$U$+$V$ methods. 
Table~\ref{table01} provides lattice constants and atomic configurations of pentahexoctite lattices relaxed by DFT-GGA and DFT+$U$+$V$ methods~\cite{PhysRevB.104.104313.Wooil.Yang, PhysRevB.102.235159.Timrov}. 
The inclusion of $U$+$V$ corrections results in decrease in $a_{1}$ from 5.788 {\AA} to 5.740 {\AA}, and increase in $a_{2}$ from 3.828 {\AA} to 3.863 {\AA}.
The self-consistent calculations of onsite Hubbard interactions ($U$) in the DFT+$U$+$V$ method provide $U_{i=1,8}=6.02$ eV, $U_{i=2,3,6,7}=6.10$ eV, and $U_{i=4,5}=5.90$ eV, where $i$ indicates carbon atom sites in Fig.~\ref{fig:configuration}. Inter-site Hubbard terms $V_{ij}$ between $i$ and $j$ sites are $V_{12}=V_{13}=3.03$ eV, $V_{24}=V_{35}=3.16$ eV, $V_{45}=3.11$ eV, $V_{18}=3.22$ eV, and $V_{23}=V_{67}=3.17$ eV. 
Note that onsite and inter-site Coulomb interactions computed in this work are comparable to those calculated for other carbon allotropes such as graphene, graphite, and BPN~\cite{PhysRevLett.106.236805, doi:10.1021/acs.nanolett.2c00528.Son}.

Next, we compute band structures of monolayer pentahexoctite by using DFT-GGA and DFT+$U$+$V$ calculations, as illustrated in Fig.~\ref{fig:bands}.
Both DFT-GGA and DFT+$U$+$V$ calculations predict that monolayer pentahexoctite is metallic, revealing nearly flat bands along $\overline{\Gamma X}$ and band crossing points near $X$. 
The computed DFT-GGA band structures are in agreement with those reported in Ref.~\onlinecite{Sharma2014}. 

Note that the term "flat band" has been used to describe a dispersionless band over the whole Brillouin zone (BZ) in the literature, for example, theoretical works mathematically formulating flat bands~\cite{PhysRevB.78.125104, PhysRevB.99.045107}, and other studies on Kagome metals~\cite{Kang2020_1} and MATBG~\cite{doi:10.1073/pnas.1108174108, PhysRevLett.122.106405, PhysRevLett.106.126802}. 
On the other hand, some references have defined a dispersionless band only along a specific symmetric line as a "flat band" as in our case~\cite{doi:10.1021/acs.jpclett.0c03816, D2CP05316K}. 
Considering these cases, we clarify that we refer to the dispersionless band along $\overline{\Gamma X}$ as the flat band throughout our work. 
In the next section, we will discuss how the dispersionless band along $\overline{\Gamma X}$ can be comprehended using the same theory applied to flat bands spanning the entire BZ.

Due to the reduced dimensionality compared to the 3D bulk systems,  
the Coulomb interactions in low-dimensional systems are likely to be less screened out. 
Therefore, it is desirable to use theoretical approaches that explicitly incorporate the Coulomb interactions rather than relying on the mean-field approximation like the GGA functionals.  
The DFT+$U$+$V$ method~\cite{Leiria_Campo_2010, PhysRevB.98.085127, PhysRevX.5.011006, PhysRevResearch.2.043410, PhysRevB.102.155117}, including on-site and inter-site Hubbard interactions, yields distinct band structures compared to the DFT-GGA calculations.
The inclusion of Hubbard interactions moves unoccupied bands up, but it shifts down occupied bands. 
The topmost partially occupied band of the DFT+$U$+$V$ method moves up by about 1-2 eV, compared with the DFT-GGA. 
In contrast, two bands below the topmost partially occupied band shift downward. 
Consequently, the nearly flat band along $\overline{\Gamma X}$, unoccupied in the DFT-GGA calculation, becomes occupied as it shifts down in the DFT+$U$+$V$ calculation. 
Similarly, band crossing points near $X$ are positioned above and below the Fermi energy $E_{F}$ in DFT-GGA and DFT+$U$+$V$ band structures, respectively. 

Fermi surfaces reflect changes in the positions of partially occupied bands relative to $E_{F}$ due to Hubbard interactions.
The DFT-GGA bands have one electron pocket around $Y$ and one hole pocket around $X$ at $E_{F}$ as shown in Fig.~\ref{fig:bands}(b). 
In addition, two open Fermi surface lines, which are mirror-symmetric to each other, traverse the BZ along the $k_{x}$ direction. 
As illustrated in Fig.~\ref{fig:bands}(c), The DFT+$U$+$V$ calculation exhibits one electron pocket around $Y$ and one hole pocket around $X$ at $E_{F}$ like the DFT-GGA calculation, but the DFT+$U$+$V$ electron (hole) pocket is smaller (larger) than that of the DFT-GGA bands. 
Apart from the electron and hole pockets, the DFT+$U$+$V$ bands has no open Fermi surface crossing the entire BZ. 

In Fig.~\ref{fig:bands}(c), it is noticeable that pristine pentahexoctite is a semimetal with a small indirect overlap between the hole pocket at $X$ and the electron pocket at $Y$. 
Considering theoretical predictions~\cite{doi:10.1080/14786436108243318, PhysRev.158.462} of the excitonic insulator phase in small indirect overlap semimetals, the semimetallic feature suggests the potential for the excitonic insulating phase in pristine monolayer pentahexoctite. 
Exciton physics can be explored by solving the Bethe-Salpeter equation~\cite{PhysRev.84.1232} with the $GW$ approximation~\cite{PhysRev.139.A796}. 
Notably, the DFT+$U$+$V$ method utilized in this study can be readily integrated with the Bethe-Salpeter equation solver in $GW$ calculation packages.

For electronic structure calculations, we consider a free-standing monolayer of pentahexoctite without the influence of a substrate. 
However, when the monolayer is placed on top of the substrate, the dielectric constant of the substrate can alter the screened Coulomb interactions, thereby impacting the band structures of pentahexoctite. 
The very weak (large) screening effect of the substrate can lead to band structures obtained from the DFT+$U$+$V$ (DFT-GGA) calculations in this work, respectively.
For intermediate substrate screening, monolayer pentahexoctite may exhibit band structures that interpolate between the results obtained from DFT-GGA and DFT+$U$+$V$ methods. 
Specifically, the overall gap between unoccupied and occupied bands may decrease in comparison with the DFT+$U$+$V$ case. 
The energy difference of two partially occupied bands at $X$ could also diminish. 
Moreover, considering that flat bands from DFT-GGA and DFT+$U$+$V$ calculations on $\Gamma X$ respectively lie above and below $E_{F}$, it is possible that the flat band may align with $E_{F}$ for monolayer pentahexoctite on top of a substrate with an intermediate dielectric constant. 

Figures~\ref{fig:bands}(d) and (e) illustrate Fermi surfaces of DFT-GGA and DFT+$U$+$V$ band structures when the Fermi level is aligned to the Dirac point. 
In the DFT-GGA case, as the Dirac point is higher than the minimum of the topmost partially occupied band, the electron pocket around $Y$ persists when the Fermi level aligns with the Dirac point. 
In contrast, for the DFT+$U$+$V$ band structures, the minimum of the topmost partially occupied band is slightly higher than the Dirac point. 
Thus there is no electron pocket around $Y$, as depicted in Fig.~\ref{fig:bands}(c). 
The Dirac point, a band crossing between the nearly flat band and the disperse band around $X$, constitutes a part of Fermi surfaces in both DFT-GGA and DFT+$U$+$V$ band structures, as shown in Figs.~\ref{fig:bands}(f) and (g), respectively. 
This implies that the point corresponds to a type-II Dirac point~\cite{Nature.527.495.Soluyanov2015, PhysRevLett.115.265304}.
We delineate electron and hole sides in the vicinity of the band crossing point by using red and blue colors in Figs.~\ref{fig:bands}(f) and (g), which well matches the typical type-II Dirac bands in the reference~\cite{Nature.527.495.Soluyanov2015}.
We also present 3D visualizations of the band surfaces touched at the type-II Dirac point, utilizing DFT-GGA band structures, as shown Figs.~\ref{fig:TypeII}(e) and (f). 
These band surfaces near the type-II Dirac point are the same as those of the type-II Dirac model~\cite{Nature.527.495.Soluyanov2015}.
The hole pocket at $X$ comprises contributions from the upper and the lower bands constituting the type-II Dirac point, as depicted in Fig.~\ref{fig:TypeII}. 
Specifically, as illustrated in Fig.~\ref{fig:TypeII}, the outer (inner) Fermi surface of the hole pocket arises from the upper (lower) band around the Dirac point. 
This result holds true for the DFT+$U$+$V$ case as well.

\begin{figure*}[t]
\begin{center}
\includegraphics[width=1.0\textwidth, clip=true]{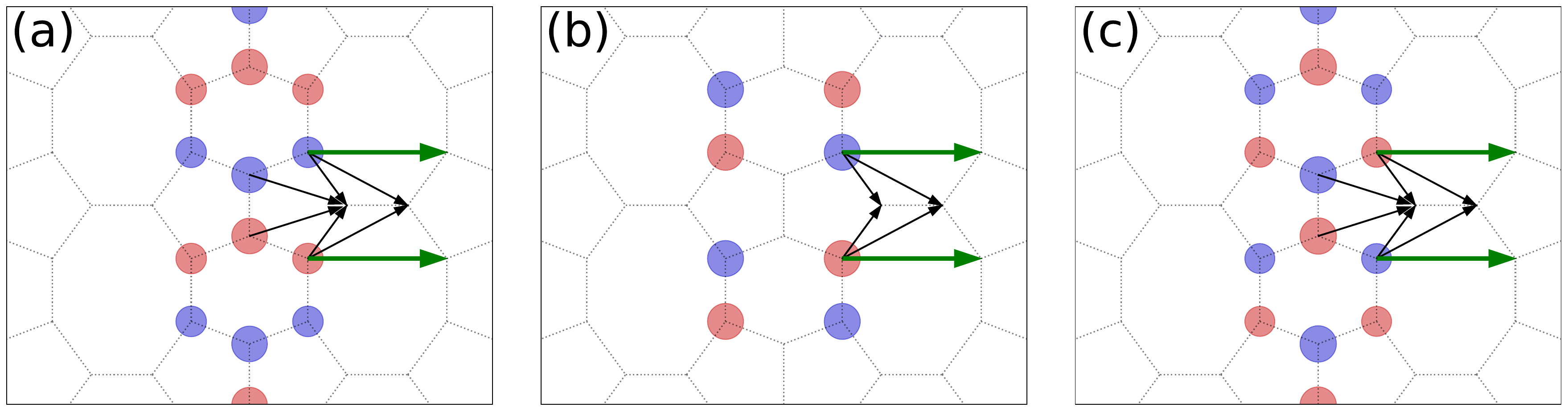} 
\end{center}
\caption{\label{fig:CLS} (Color online) Compact localized states associated with (a) the lowest, (b) the second lowest, and (c) the third lowest flat bands on $\overline{\Gamma X}$. The lowest flat band constitutes the type-II Dirac bands, and the second and the third flat bands correspond to the sixth and the eighth bands of Wannier Hamiltonians as shown in Fig.~\ref{fig:truncation}. Amplitudes of compact localized states are denoted by circles, whose radii are proportional to the amplitudes. Red and blue colors represent the opposite signs. Black thin arrows represent hopping processes contritubing to destructive interferences. Green thick arrows are non-destructive hopping processes leading to non-zero dispersion.}
\end{figure*}

\begin{figure}[t]
\begin{center}
\includegraphics[width=1.0\columnwidth, clip=true]{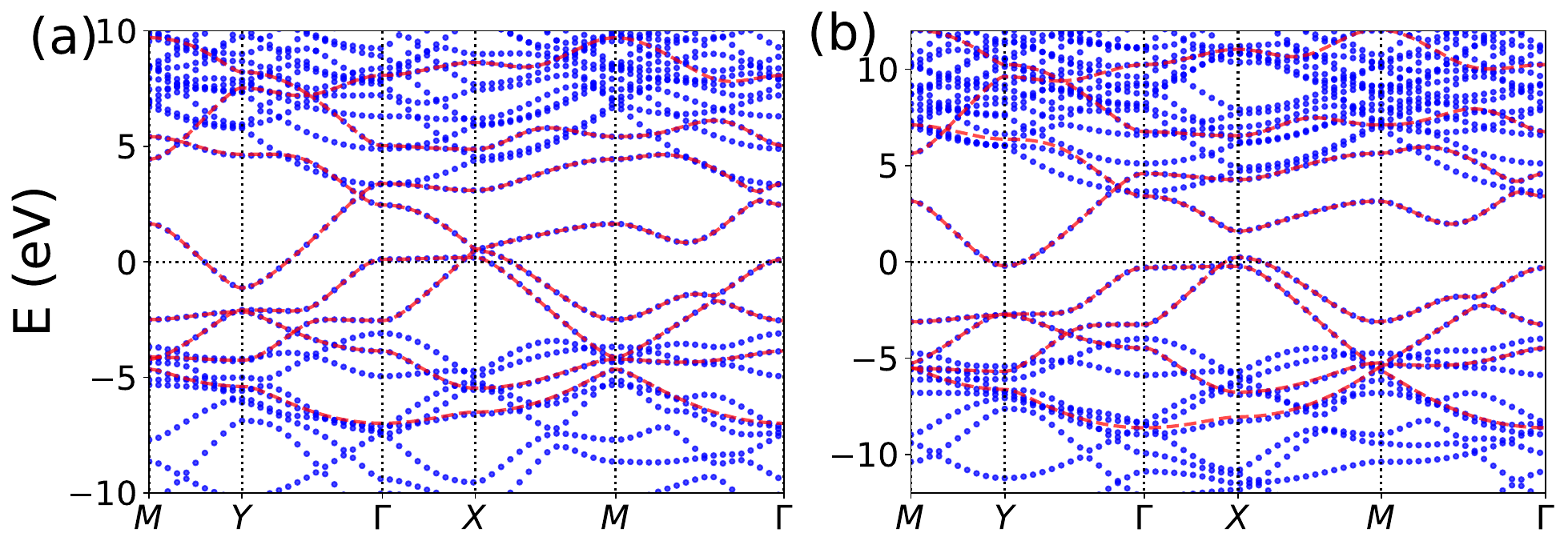} 
\end{center}
\caption{\label{fig:Wannier} (Color online) Comparison between first-principles electronic bands (blue circles) and Wannier-interpolated ones (red solid lines) calculated by using (a) DFT-GGA and (b) DFT+$U$+$V$ methods.}
\end{figure}

\begin{figure*}[t]
\begin{center}
\includegraphics[width=1.0\textwidth, clip=true]{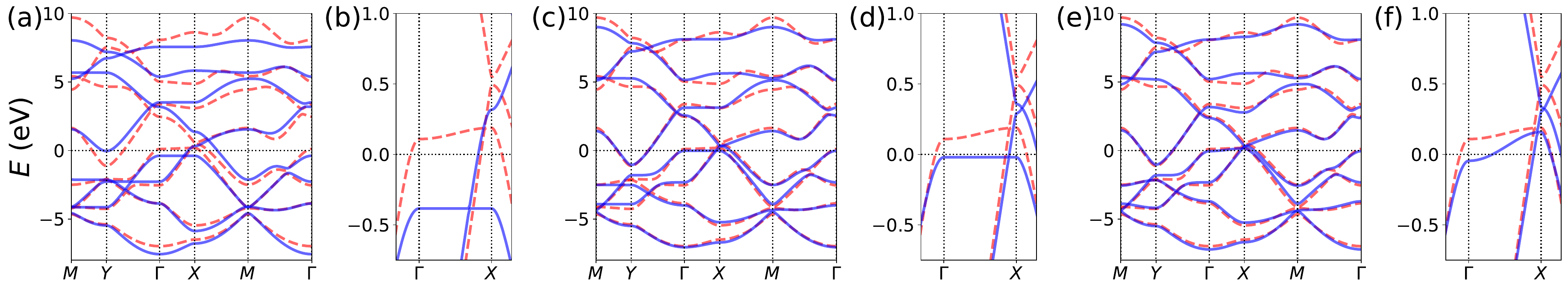} 
\end{center}
\caption{\label{fig:truncation} (Color online) Band structures (blue solid lines) calculated by Wannier Hamiltonians whose hopping parameters are truncated when distances between Wannier functions are beyond cutoff radii (a) 1.5 \AA, (c) 2.8 \AA,  and (e) 3.2 \AA. Flat bands on $\overline{\Gamma X}$ corresponding to (a), (b), and (c) are magnified at (b), (d), and (f), respectively. The reference band structures computed by DFT-GGA calculations are denoted by red-dashed lines.}
\end{figure*}

\begin{figure*}[t]
\begin{center}
\includegraphics[width=1.0\textwidth, clip=true]{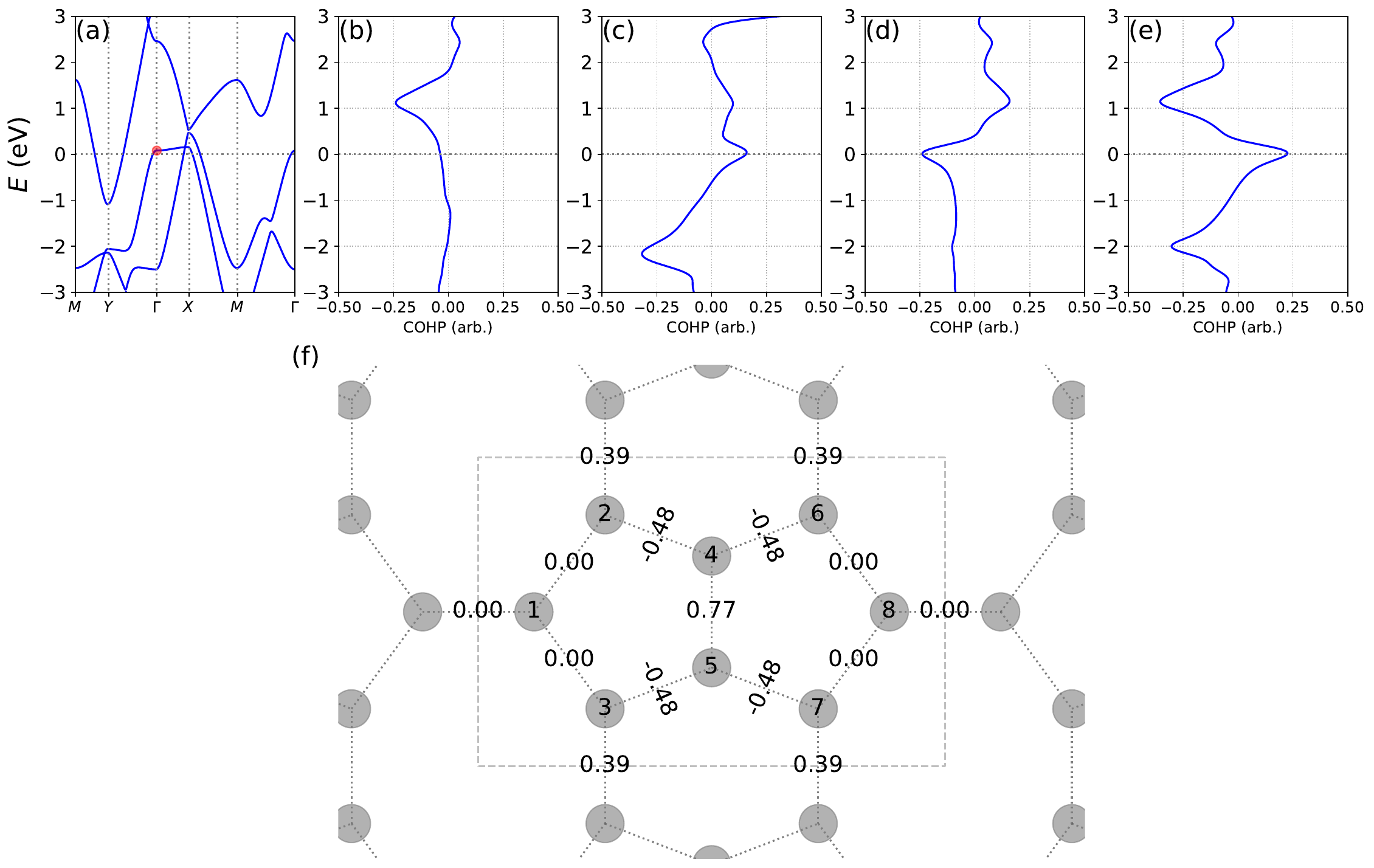} 
\end{center}
\caption{\label{fig:cohp} (Color online) Crystal orbital Hamiltonian population (COHP) analysis of the flat band. (a) The DFT-GGA band structure around $E_{F}$. The Fermi level $E_{F}=0$ is denoted by black dotted lines. COHP curves between carbon atoms (b) 1 and 2, (c) 2 and 3, (d) 2 and 4, and (e) 4 and 5. Negative, positive, and zero values of COHP represent bonding, antibonding, and non-bonding characters, respectively. (f) $k$-resolved, band-resolved COHP for the flat band at $\Gamma$, which is denoted by red circle in (a). COHP values are indicated on bonds between carbon atoms.}
\end{figure*}

\section{ANALYSIS OF FLAT BANDS}
Here we investigate the origin of flat bands along $\overline{\Gamma X}$. 
In the electronic structure theory, band dispersion $E_{n}(\mathbf{k})$ is linked to the electron velocity as follows:
\begin{equation}
\mathbf{v}_{n}(\mathbf{k}) = \frac{1}{\hbar}\frac{\partial}{\partial \mathbf{k}}E_{n}(\mathbf{k}) - \frac{e}{\hbar}\mathbf{E} \times \Omega_{n}(\mathbf{k}),
\end{equation}
where $e$ is the electron charge, $\mathbf{E}$ is an electric field, and $\Omega_{n}(\mathbf{k})$ is the Berry curvature of the $n$th band. 
In the absence of an electric field, velocity is determined by the usual band dispersion. 
It means that a disperionless band corresponds to an electron state with zero velocity.
It is suggested that flat bands can be understood in terms of compact localized states (CLS), which are strictly confined inside a spatially finite region but have zero amplitude outside the region~\cite{PhysRevB.78.125104, PhysRevB.99.045107}. 
The presence of compact localized states arises from destructive interference. 
If quantum-mechanical hopping processes from a state to every point outside the state are canceled out via destructive interference, electrons in the state cannot escape and remain confined. 
Since electrons in compact localized states cannot transfer outside, their band dispersion becomes flat, exhibiting zero velocity. 

The concept of the compact localized state and destructive interference is applicable even when a flat band forms only along a specific direction, as is the case of our interest. 
Illustrated in Fig.~\ref{fig:CLS}, compact localized states corresponding to flat bands along $\overline{\Gamma X}$ are strictly confined along the $\hat{x}$ direction, but extend along the $\hat{y}$ direction. 
Hopping processes out of compact localized states along the $\hat{x}$ direction are canceled out by destructive interference. 
We provide a detailed analysis of compact localized states of monolayer pentahexoctite in the following discussion. 

To investigate the aforementioned destructive interference of compact localized states, we first construct a tight-binding model by using maximally localized Wannier functions~\cite{RevModPhys_84_1419_2012_Marzari}. 
Since electronic bands around $E_{F}$, including partially occupied bands of our interest, primarily originate from $p_{z}$ orbitals of carbon atoms, we build the Wannier Hamiltonian based on $p_{z}$-like maximally localized Wannier functions as follows:
\begin{equation}
\mathcal{H}_{ij}(\mathbf{k}) = \sum_{\mathbf{R}}e^{-i\mathbf{k}\cdot\mathbf{R}_{ij}}\langle \phi_{i}(\mathbf{r}_{i}+\mathbf{R})\left|\mathcal{H}\right|\phi_{j}(\mathbf{r}_{j})\rangle,    
\end{equation}
where $\{\mathbf{R}\}$ represents lattice vectors, $\mathbf{r}_{i}$ is the position vector of the $i$th carbon atom inside the unit cell, 
$\mathbf{R}_{ij} \equiv \mathbf{R}+\mathbf{r}_{i} - \mathbf{r}_{j}$
and $|\phi(\mathbf{r}_{i}+\mathbf{R})\rangle$ is the $p_{z}$ orbital of $i$th carbon atom located at the lattice vector $\mathbf{R}$. 

Figure~\ref{fig:Wannier} illustrates electronic band structures interpolated by $p_{z}$-type Wannier functions, compared with first-principles calculations based on DFT-GGA and DFT+$U$+$V$ methods. 
In both cases, the Wannier-interpolated bands well reproduce the nearly flat band around the Fermi level. 
Although the Wannier Hamiltonians corresponding to DFT-GGA and DFT+$U$+$V$ band structures have different onsite energies of $p_{z}$-type Wannier functions and hopping parameters, the origin of the nearly flat band along $\overline{\Gamma X}$ can be explained by the same theory involving the formation of compact localized states due to destructive interference. 
Therefore, we use the Wannier Hamiltonian interpolating DFT-GGA band structures in the following investigation. 

We construct Wannier-based tight-binding models by keeping hopping parameters only between Wannier functions whose inter-distances are within a specific cutoff radius $R_{c}$. 
The truncated Wannier Hamiltonian is expressed as follows:
\begin{equation}
\mathcal{H}_{ij}(\mathbf{k}) = \sum_{\mathbf{R} \atop |\mathbf{R}_{ij}| \leq R_{c}}e^{-i\mathbf{k}\cdot\mathbf{R}_{ij}}\langle \phi_{i}(\mathbf{r}_{i}+\mathbf{R})\left|\mathcal{H}\right|\phi_{j}(\mathbf{r}_{j})\rangle,    
\end{equation}
where the summation over $\mathbf{R}$ is restricted to the case where $|\mathbf{R}_{ij}|\leq R_{c}$. 
We investigate the evolution of the flat band by increasing the cutoff radius. 

First, with a cutoff radius of 1.5 \AA, the truncated Wannier Hamiltonian represents a tight-binding model encompassing solely nearest neighbor interactions. 
The resulting band structures are depicted in Figs.~\ref{fig:truncation}(a) and (b). 
While DFT band structures have a nearly flat band with a small but finite slope along $\overline{\Gamma X}$, the nearest-neighbor tight-binding model exhibits a perfectly flat band around $E_{F}$. 
Furthermore, two perfect flat bands on $\overline{\Gamma X}$ emerge in the sixth and the eighth bands of the nearest-neighbor tight-binding model, as shown in Fig.~\ref{fig:truncation}. 
Extending the truncation radius $R_{c}$ to 2.8 {\AA} incorporates second nearest-neighbor interactions into the truncated Wannier Hamiltonian. 
As shown in Figs.~\ref{fig:truncation}(c) and (d), the inclusion of the second nearest neighbors does not affect the perfect flatness of the three bands on $\overline{\Gamma X}$ observed in the nearest-neighbor tight-binding model. 

Figure~\ref{fig:CLS} illustrates the three compact localized states on $\overline{\Gamma X}$ and their destructive interference paths contributing to the perfect flatness.  
Denoting mirror-reflection symmetries with respect to $yz$ and $zx$ planes as $\mathcal{M}_{x}$ and $\mathcal{M}_{y}$ respectively, the compact localized states of the three flat bands are all anti-symmetric with respect to $\mathcal{M}_{y}$. 
While the compact localized states of the lowest and third lowest flat bands are symmetric with respect to $\mathcal{M}_{x}$, the compact localized state of the second lowest flat band is anti-symmetric with respect to $\mathcal{M}_{x}$. 
Defining the hopping parameter as 
\begin{equation}
t_{ij}(\mathbf{R}) \equiv \langle \phi_{i}(\mathbf{r}_{i}+\mathbf{R})\left|\mathcal{H}\right|\phi_{j}(\mathbf{r}_{j})\rangle,
\end{equation}
the Wannier Hamiltonian gives the nearest neighbor hopping parameters $t_{86}(\mathbf{0})=t_{87}(\mathbf{0})=-2.56$ eV, and the second nearest neighbor ones $t_{84}(\mathbf{0})=t_{85}(\mathbf{0})=-0.0879$ eV and $t_{16}(\mathbf{a}_{1})=t_{17}(\mathbf{a}_{1})=0.211$ eV. 
This implies that the tight-binding model is symmetric with respect to $\mathcal{M}_{y}$. 
Consequently, since all compact localized states are anti-symmetric with respect to $\mathcal{M}_{y}$, all hopping processes out of the compact localized states are exactly canceled out, leading to destructive interference. 
Explicitly, for the compact localized state
\begin{equation}
|\textrm{CLS}\rangle=\sum_{i=1}^{8}\sum_{n=-\infty}^{\infty} \alpha_{i}|\phi_{i}(\mathbf{r}_{i} + \mathbf{n\mathbf{a}_{2}})\rangle, 
\end{equation}
the sum of paths up to the second nearest neighbor hopping vanishes:
\begin{eqnarray}
\sum\textrm{(path)}&=&\alpha_{6}t_{86}(\mathbf{0})+\alpha_{7}t_{87}(\mathbf{0})+\alpha_{4}t_{84}(\mathbf{0}) \nonumber\\
&&+\alpha_{5}t_{85}(\mathbf{0})+\alpha_{6}t_{16}(\mathbf{a}_{1}) + \alpha_{7}t_{87}(\mathbf{a}_{1})=0,\nonumber\\
\end{eqnarray}
where $\alpha_{4}=-\alpha_{5}$ and $\alpha_{6}=-\alpha_{7}$.

If the truncation cutoff is extended to include third nearest neighbor hopping processes in the tight-binding model, the perfect dispersionless nature of the three flat bands is no longer maintained. 
For instance, when $R_{c}$ is 3.2 \AA, the three flat bands exhibit small but finite dispersion, as shown in Figs.~\ref{fig:truncation}(e) and (f). 
In this case, the third nearest neighbor hopping terms $t_{26}(\mathbf{a}_{1})=t_{27}(\mathbf{a}_{1})=-0.200$ eV are included, as illustrated by the green thick arrows in Figs.~\ref{fig:CLS}(a)-(c).
These hopping contributions are not canceled out by other hopping processes through destructive interference. 
Hence these contributions introduce small but finite dispersion to the three bands on $\overline{\Gamma X}$ that are perfectly flat up to the second nearest neighbor interactions. 
We remark that in the DFT calculation or the full Wannier Hamiltonian, the nearly flat band around $E_{F}$ possesses a bandwidth of approximately 80 meV.  

We also performed the crystal orbital Hamiltonian population (COHP) analysis~\cite{doi:10.1021/j100135a014} to further characterize the flat band. 
Figures~\ref{fig:cohp}(b)-(e) illustrate energy-resolved COHP curves between carbon sites 1-2, 2-3, 2-4, and 4-5. Around $E_{F}$, the 2-4 bond exhibits negative COHP, while the 2-3 and 4-5 bonds display positive COHP. 
Considering that negative and positive values of COHP indicate bonding and anti-bonding characters respectively, the 2-4 bond corresponds to bonding interactions, whereas the 2-3 and 4-5 bonds represent anti-bonding interactions. 
The 1-2 bond has a very small negative COHP, implying that the 1-2 bond is nearly non-bonding. 
Furthermore, we computed the $k$-resolved, band-resolved COHP of the flat band at $\Gamma$, as depicted in Fig.~\ref{fig:cohp}(f).
The 2-4, 3-5, 4-6, and 5-7 bonds, which are equivalent to one another via mirror symmetries $\mathcal{M}_{x}$ and $\mathcal{M}_{y}$, exhibit bonding interactions with negative COHP. 
In contrast, the 2-4, 4-5, and 6-7 bonds are identified as anti-bonding interactions with positive COHP.
The 1-2, 1-3, 6-8, and 7-8 bonds have zero COHP, indicating non-bonding interactions. 
These COHP values and corresponding bonding characters are consistent with the compact localized state shown in Fig.~\ref{fig:CLS}(a).

\section{MAGNETIC ORDERING}
The Stoner theory of magnetism for itinerant electrons~\cite{Stoner1938} provides the condition for the magnetic transition known as the Stoner criterion, with $ID(E_{F})>1$, where $I$ is the Stoner factor determined by the exchange-correlation interaction, and $D(E_{F})$ is the density of states at $E_{F}$. 
Since the flat band possesses a significantly large density of states, the Stoner criterion suggests the possibility of magnetic ordering in monolayer pentahexoctite if the flat band is located around $E_{F}$.
Hole doping is one possible mechanism to position the flat band around $E_{F}$.  
By depopulating electrons with hole-doping methods, the flat band can relatively move up to $E_{F}$, so it is expected that monolayer pentahexoctite can exhibit magnetic ordering according to the Stoner criterion.  
Here, we explore the possibility of the magnetic transition in monolayer pentahexoctite by controlling hole-doping concentrations.

As shown in Fig.~\ref{fig:bands}(a), DFT+$U$+$V$ band structures are spin-degenerate, and the nearly flat band is located about 0.3 eV below $E_{F}$. 
To adjust the position of the flat band, we vary hole doping concentrations $n_{h}$ up to 0.40 hole/unit cell. 
It is noteworthy that we control hole doping concentrations by changing the total charge of the system with a compensating jellium background as implemented in \textsc{Quantum-Espresso}~\cite{JPhys_CM_29_465901_2017}. 
This hole-doping simulation corresponds to the experimental hole-doping method based on an electrostatic dual-gated device structure~\cite{PhysRevLett.105.166601}.  
This doping method preserves the chemical components and atomic
structures of the target material, allowing us to maintain monolayer pentahexoctite in its pristine state.
Hence, this hole doping method is suitable for our purpose. 
The doping with real atoms can be adopted to manipulate hole concentrations. 
However, it is important to recognize that this doping method introduces changes beyond electron/hole concentrations, including alternations in atomic structures and broken lattice symmetries, since dopants act as impurities. 
Thus, this could lead to defect-induced magnetism as observed in defected graphene~\cite{PhysRevLett.93.187202, doi:10.1021/nl9020733, PhysRevLett.91.017202}.
In the Supporting Information, we discuss the magnetic ordering of pentahexoctite doped with boron atoms, presenting it as an illustrative example of doping with real atoms.

\begin{figure}[H]
\begin{center}
\includegraphics[width=1.0\columnwidth, clip=true]{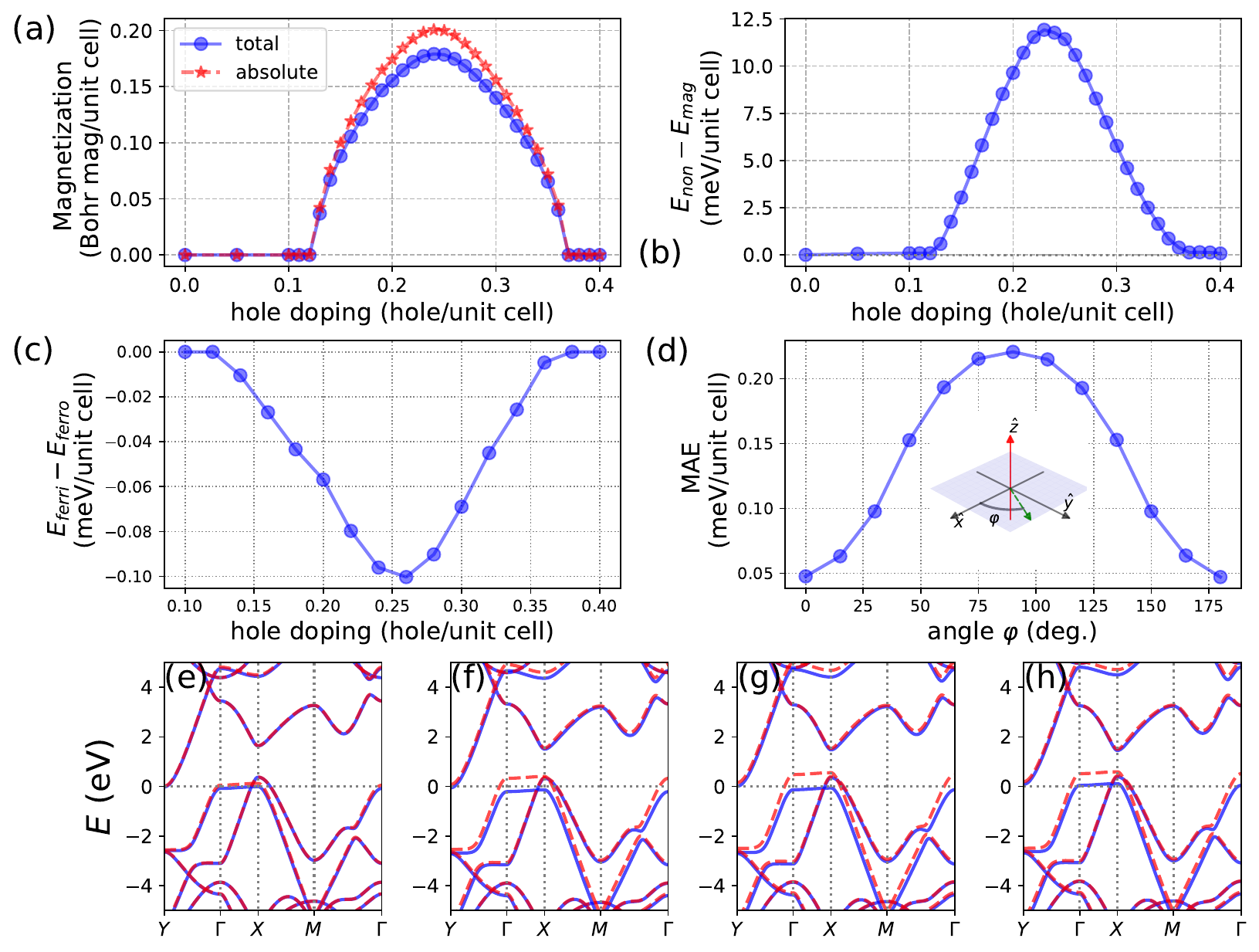} 
\end{center}
\caption{\label{fig:magnetization} (Color online) (a) Magnetization of single-layer pentahexoctite as a function of hole doping concentration. Red dashed lines with star markers (blue solid lines with close circles) denote absolute (total) magnetization. (b) Total energy differences $E_{non} - E_{mag}$ between non-magnetic and magnetic phases. (c) Total energy differences $E_{ferri} - E_{ferro}$ between ferrimagnetic and ferromagnetic phases. (d) Magnetic anisotropic energy (MAE) $E_{\parallel}(\varphi)-E_{\perp}$ when $n_{h}=0.25$ as a function of azimuthal angle $\varphi$ representing in-plane magnetization direction. The inset is a schematic illustration of azimuthal angle $\varphi$. Band structures at hole doping concentrations (e) 0.13 (f) 0.20 (g) 0.25 (h) 0.30 in unit of hole per unit cell. Energy bands of the magnetic state are drawn by blue solid lines and red dashed ones, which correspond to opposite spin orientations. Here $E_{F}$ is set to be zero.}
\end{figure}

\begin{figure}[H]
\begin{center}
\includegraphics[width=1.0\columnwidth, clip=true]{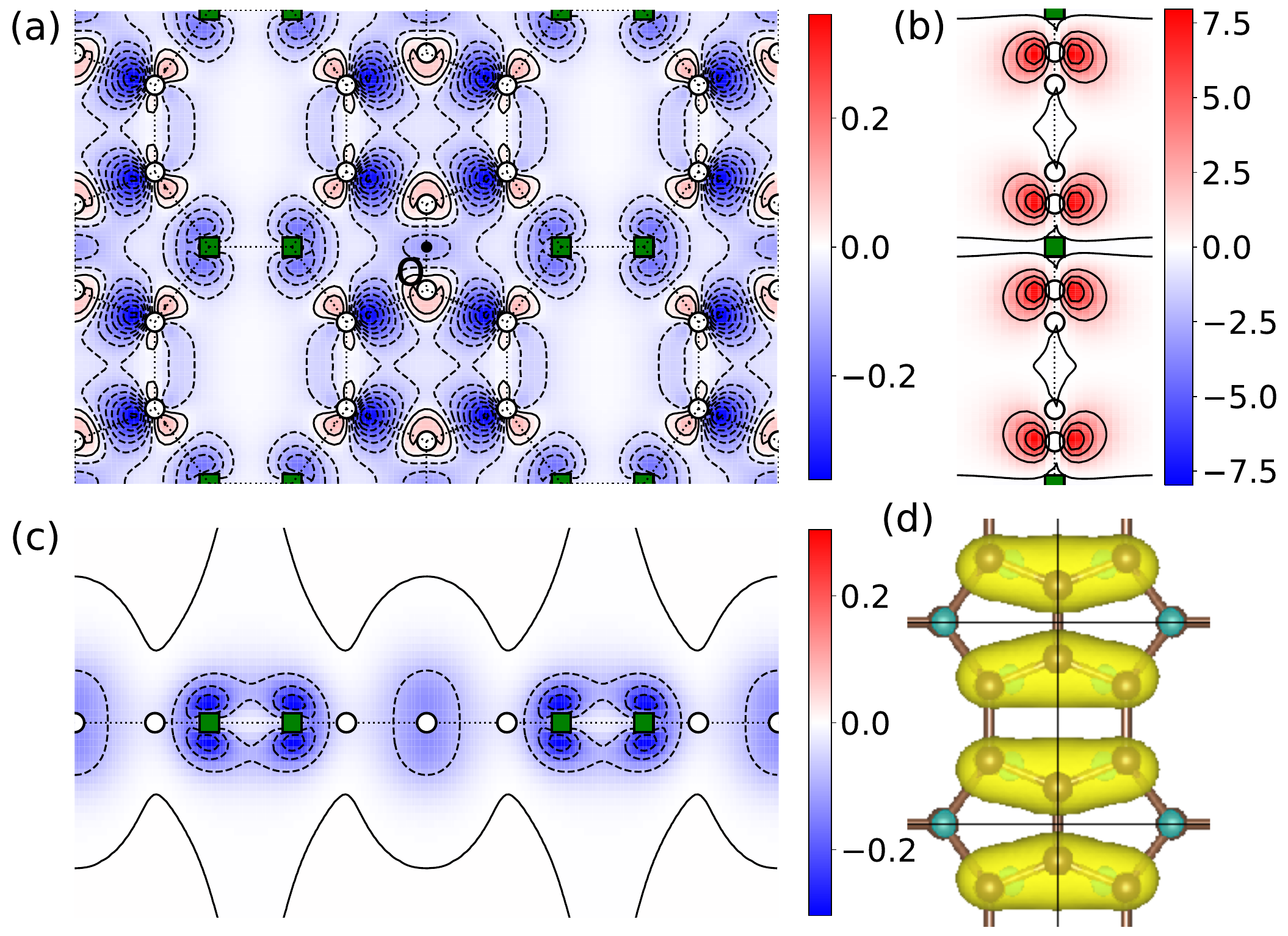} 
\end{center}
\caption{\label{fig:spin_density} (Color online) Contour plots of spin polarization density $\Delta \rho=\rho_{\uparrow} - \rho_{\downarrow}$ on (a) $xy$, (b) $yz$, and (c) $zx$ planes including the origin $O$, when the hole concentration $n_{h}$ is 0.24 hole per unit cell. Spin polarization density is rescaled for clarity. The same rescaling factor is used for (a), (b), and (c). Red color and solid contour lines (blue color and dashed contour lines) represent positive (negative) spin polarization density. Green squares and white circles represent carbon atoms 1 and 8, and atoms from 2 to 7, respectively. (d) the top view of the spin polarization density isosurface when $\Delta \rho=2.14\times10^{-3}$ $\AA^{-3}$. Yellow and green isosurfaces represent majority and minority spins, respectively.}
\end{figure}

Figure~\ref{fig:magnetization}(a) shows absolute and total magnetizations as functions of hole doping concentrations $n_{h}$. 
Monolayer pentahexoctite starts to exhibit magnetism at $n_{h} = 0.12$ hole per unit cell. 
A comparison between absolute and total magnetizations indicates that the total magnetization is slightly smaller than the absolute one, suggesting ferrimagnetic behavior of monolayer pentahexoctite. 
The ferrimagnetic phase persists up to $n_{h} = 0.37$ hole per unit cell. 
Band structures corresponding to the ferrimagnetic phase for selected hole doping concentrations $n_h$ = 0.13, 0.20, 0.25 and 0.30 hole per unit cell are calculated using the DFT+$U$+$V$ method, as illustrated in Figs.~\ref{fig:magnetization}(e)-(h). 
In the ferrimagnetic phase, spin-up and spin-down bands are no longer degenerate, with the most evident splitting observed at nearly flat bands around $E_{F}$.
This splitting is enhanced as the induced magnetization increases.

We computed the total energy difference $\Delta E = E_{non} - E_{mag}$, where $E_{non}$ and $E_{mag}$ respectively represent total energies of the non-magnetic and the magnetic phases, as shown in Fig.~\ref{fig:magnetization}(b). 
The total energy difference $\Delta E$ reaches the maximum value, approximately 12 meV, at $n_{h}=0.25$ hole/unit cell. 
Using the mean-field approximation $k_{B} T_{C}\approx \frac{2}{3}\Delta E$~\cite{RevModPhys.82.1633}, this total energy difference corresponds to the estimated Curie temperature $T_{C}$, about $93$ K. 
Considering the thermal fluctuation of spins, the Curie temperature could be lower than the mean-field estimation~\cite{Moriya1985}.

Additionally, we calculated the magnetic anisotropic energy (MAE) of the system, defined as $E_{\textrm{MAE}}(\varphi) = E_{\parallel}(\varphi)-E_{\perp}$, where $E_{\parallel}(\varphi)$ is the in-plane energy when the in-plane magnetization axis has the azimuthal angle $\varphi$ as shown in the inset of Fig.~\ref{fig:magnetization}(d), and $E_{\perp}(\varphi)$ is the out-of-plane energy with magnetization aligned along the $\hat{z}$ axis. 
Figure~\ref{fig:magnetization}(d) presents $E_{\textrm{MAE}}(\varphi)$ as a function of the azimuthal angle $\varphi$, ranging from about 0.050 to 0.221 meV/unit cell. 
When the in-plane magnetization aligns with the $\hat{y}$ direction ($\varphi=\pi/2$), $E_{\textrm{MAE}}$ reaches its maximum value of 0.221 meV/unit cell. 
It is worth noting that the magnetic anisotropy could be enhanced by using external fields or substrates.~\cite{Gong2017, doi:10.1080/23746149.2018.1432415}

Figure~\ref{fig:magnetization}(c) shows the total energy difference $E_{\textrm{ferri}}-E_{\textrm{ferro}}$ between the ferrimagnetic ordering and the ferromagnetic ordering as a function of hole doping concentrations. 
Here, we determined the ferromagnetic ordering energy $E_{\textrm{ferro}}$ by performing constrained magnetization calculations~\cite{C2DT31662E}, where the magnitude of magnetic moments is preserved, but all magnetic moments are aligned in the same direction.  
As shown in Fig.~\ref{fig:magnetization}(c), the transition from the ferrimagnetic phase to the ferromagnetic one requires a positive energy cost, whose maximum is about 0.10 meV at $n_{h}\approx 0.25$ hole/unit cell.

To visualize the ferrimagnetic ordering, we calculate spin polarization density $\Delta \rho = \rho_{\uparrow} - \rho_{\downarrow}$, representing the difference between spin-up density $\rho_{\uparrow}$ and spin-down density $\rho_{\downarrow}$. 
Figures~\ref{fig:spin_density}(a)-(c) illustrate spin polarization densities on $xy$, $yz$, $zx$ planes containing the origin $O$ located at the middle of atoms 4 and 5, respectively. 
The three-dimensional isosurface of the spin polarization density is also depicted in Fig.~\ref{fig:spin_density}(d). 
The majority spin density concentrates on atoms from 2 to 7, while the minority spin density is located at atoms 1 and 8 as well as interstitial sites between 2 and 4, 4 and 6, 3 and 5, and 5 and 7.  

\begin{figure}[t!]
\begin{center}
\includegraphics[width=0.8\columnwidth, clip=true]{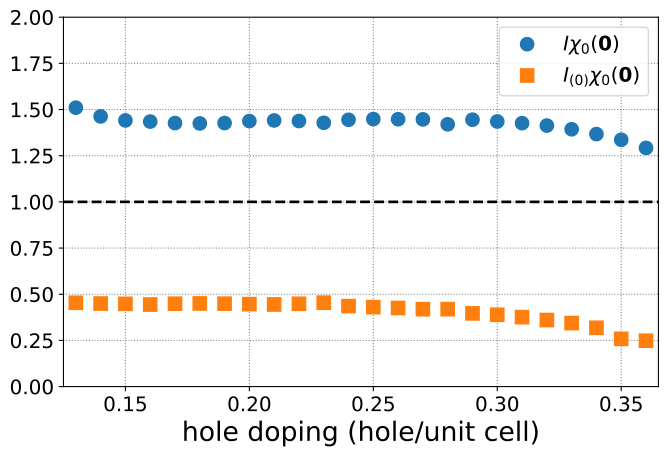} 
\end{center}
\caption{\label{fig:stoner} (Color online) (a) $I_{(0)}\chi_{0}\left(\mathbf{0}\right)$ (orange square) and $I\chi_{0}\left(\mathbf{0}\right)$ (blue circle) for DFT-GGA and DFT+$U$+$V$ calculations, respectively, for hole doping concentrations at which ferrimagnetic transition occurs. The Stoner criterion for magnetic instability is denoted by a black dashed line.}
\end{figure}
To verify that the magnetic state of hole-doped pentahexoctite satisfies the Stoner criterion, we compute the spin susceptibility by using the random phase approximation (RPA)~\cite{DAS2021110758}:
\begin{equation}
\label{Eq:RPA}
    \chi(\mathbf{q}=\mathbf{0}) = \frac{\chi_{0}(\mathbf{0})}{1-I\chi_{0}(\mathbf{0})},
\end{equation}
where $\chi_{0}(\mathbf{0})$ is the bare one-electron susceptibility at $\mathbf{q}=\mathbf{0}$, and $I$ is the Stoner factor from the exchange-correlation interaction~\cite{DAS2021110758}. 
Here $I\chi_{0}(\mathbf{0}) \geq 1$ indicates the magnetic instability, which is the Stoner criterion for magnetic phase transition~\cite{DAS2021110758, Stoner1938}. Note that $\chi_{0}(\mathbf{0})=D(E_{F})$. 
For $\mathbf{q}=\mathbf{0}$, we calculate the spin susceptibility $\chi(\mathbf{0})$ by using the fixed spin moment method~\cite{DAS2021110758}. 
When the total energy $E_{tot}$ in DFT calculations is decomposed as 
\begin{equation}
    E_{tot}(\mathbf{M}) = \left[E_{tot}(\mathbf{M}) - E_{xc}(\mathbf{M})\right] + E_{xc}(\mathbf{M}),
\end{equation}
where $E_{xc}(\mathbf{M})$ denotes the exchange-correlation energy when the total magnetization is $\mathbf{M}$ in the fixed spin moment calculation, 
$\chi(\mathbf{0})$, $\chi_{0}(\mathbf{0})$, and $I$ are computed as follows:
\begin{eqnarray}
    \label{Eq:chi}\chi^{-1}(\mathbf{0}) &=& \frac{\delta^{2} E_{tot}}{\delta \mathbf{M}^{2}} \\
    \label{Eq:chi0}\chi_{0}^{-1}(\mathbf{0}) &=& \frac{\delta^{2}}{\delta \mathbf{M}^{2}}\left(E_{tot} - E_{xc}\right) \\
    \label{Eq:I}I &=& -\frac{\delta^{2} E_{xc}}{\delta \mathbf{M}^{2}}.
\end{eqnarray}
Thus it implies that 
\begin{eqnarray}
    \chi^{-1}(\mathbf{0}) = \chi^{-1}_{0}(\mathbf{0}) - I,
\end{eqnarray}
which is equal to Eq.~(\ref{Eq:RPA}). 
In the DFT+$U$+$V$ method, the exchange-correlation energy $E_{xc}$ is expressed as 
\begin{equation}
    E_{xc} = E_{xc}^{(0)} + E_{U+V} - E_{dc},
\end{equation}
where $E_{xc}^{(0)}$ is the exchange-correlation energy in the GGA level, $E_{U+V}$ is the Hubbard interaction energy with $U$ and $V$, and $E_{dc}$ is the double-counting term~\cite{PhysRevResearch.2.043410, PhysRevB.102.155117, PhysRevX.5.011006, Leiria_Campo_2010}. In the DFT-GGA calculations, the Stoner factor $I_{(0)}$ is 
\begin{equation}
    \label{Eq:I0}I_{(0)} = \frac{\delta^{2} E_{xc}^{(0)}}{\delta \mathbf{M}^{2}}.
\end{equation}
Using Eqs.~(\ref{Eq:chi0}), (\ref{Eq:I}), and (\ref{Eq:I0}), we compute the Stoner criteria $I_{(0)}\chi_{0}(\mathbf{0})$ and $I\chi_{0}(\mathbf{0})$ for DFT-GGA and DFT+$U$+$V$ calculations, respectively, when hole doping leads to magnetic transition. 
As seen in Fig.~\ref{fig:stoner}, DFT-GGA calculations show $I_{(0)}\chi_{0}(\mathbf{0})<1$, implying that the Stoner criterion is not satisfied in the DFT-GGA scheme. 
This is consistent with the fact that magnetic ordering is not observed in DFT-GGA calculations. 
On the other hand, the inclusion of Hubbard interactions $U$ and $V$ in DFT calculations enhances the exchange-correlation interaction, thereby leading to the magnetic instability $I\chi_{0}(\mathbf{0})>1$ as shown in Fig.~\ref{fig:stoner}.
It signifies that the explicit inclusion of Coulomb interactions is crucial to investigate the magnetic transition of pentahexoctite.

\section{\label{sec:conclusions}CONCLUSIONS}
In this study, we investigated the nearly flat bands of monolayer pentahexoctite and their electronic structure properties by using the DFT+$U$+$V$ method explicitly incorporating on-site and inter-site Hubbard interactions.
Our analysis of Fermi surfaces confirmed that monolayer pentahexoctite is the type-II Dirac material where the nearly flat band around the Fermi level intersects with a dispersive band at the Dirac point. 
This electronic behavior distinguishes pentahexoctite from graphene, which is a representative type-I Dirac material. 

We explored the physical origin of the nearly flat bands of monolayer pentahexoctite using compact localized states and quantum-mechanical destructive interference. 
Two key features contribute to the formation of compact localized states: the mirror symmetry with respect to the vertical plane parallel to the nearly flat band, and the connection of compact localized states to neighboring atomic structures through carbon dimers. 
The anti-symmetric compact localized states with respect to the mirror symmetry lead to the cancellation of all hopping processes converging to dimer bridges via destructive interference. 
These features explain the formation of a flat band in monolayer pentahexoctite.
Furthermore, it is anticipated that other all-carbon 2D materials exhibiting these features can also host flat bands. 

Considering the enhanced density of states associated with flat bands, we investigated the possibility of magnetic ordering in monolayer pentahexoctite by using extended Hubbard interactions. 
Our calculations indicated that pristine monolayer pentahexoctite is non-magnetic with the nearly flat band located below the Fermi level. 
Hole doping leads to a ferrimagnetic magnetic transition in monolayer pentahexoctite, aligning the nearly flat band with the Fermi level. 
This suggests that monolayer pentahexoctite can serve as a candidate for magnetic 2D all-carbon materials, suitable for future spintronics and nanoelectronics applications.

We make a closing remark that our investigation into flat bands and their magnetic properties extends beyond pentahexoctite. 
Among all-carbon allotropes predicted to date, one can identify 2D all-carbon crystals hosting flat bands, where the combination of crystal symmetries and structural geometries can facilitate the formation of compact localized states. 
For carbon allotropes featuring flat bands near the Fermi level, the possibility of engineering their magnetism through external perturbations, such as electron or hole doping, opens the door to discovering other candidates for pristine all-carbon 2D materials.

\section*{Acknowledgments}
We thank Sang-Hoon Lee, Wooil Yang, and Young-Woo Son for sharing the DFT+$U$+$V$ codes.
We thank Hosub Jin for the fruitful discussions.
This work was supported by the National Research Foundation of Korea (NRF) grant funded by the Korea government (Grant No.~NRF-2022R1F1A1074670) and by the Open KIAS Center at Korea Institute for Advanced Study.

\section*{AUTHOR DECLARATIONS}
\subsection*{Conflict of Interest}
The authors have no conflicts to disclose.
\subsection*{Author contributions}
\noindent \textbf{Sejoong Kim}: Conceptualization;
Data curation;
Formal analysis;
Funding acquisition;
Investigation;
Methodology;
Project administration;
Resources;
Software;
Supervision;
Validation;
Visualization;
Writing – original draft;
Writing – review \& editing;

\bibliography{bib}

\end{document}